\documentclass[12pt,letter]{article}

\usepackage{graphicx, epsfig, color,cite}
\usepackage{amsmath}
\usepackage{amssymb}
\usepackage{subfigure}

\def\delew{\Delta_{\rm EW}}

\def\to{\rightarrow}

\def\bi{\begin{itemize}}
\def\ei{\end{itemize}}

\def\sps1ap{SPS1a$^\prime$}
\def\c1p{C1$^\prime$}

\def\tst{\tilde t}

\def\tg{\tilde g}

\def\tw{\widetilde W}
\def\tz{\widetilde Z}
\def\alt{\lesssim}
\def\agt{\gtrsim}
\def\be{\begin{equation}}  
\def\ee{\end{equation}}  
\def\bea{\begin{eqnarray}}  
\def\eea{\end{eqnarray}}  
\def\beas{\begin{eqnarray*}}  
\def\eeas{\end{eqnarray*}}  
\newcommand\prd[3]{{\it Phys.\ Rev.\ }{\bf D #1} (#2) #3}

\newcommand\plb[3]{{\it Phys.\ Lett.\ }{\bf B #1} (#2) #3}
\newcommand\jhep[3]{{\it J. High Energy Phys.\ }{\bf #1} (#2) #3}

\newcommand\npb[3]{{\it Nucl.\ Phys.\ }{\bf B #1} (#2) #3}

\newcommand{\hepph}[1]{hep-ph/#1}

\begin{document}
\begin{titlepage}
\begin{flushright}
UT-16-23 \\
NSF-KITP-16-083 \\
UMN-TH-3532/16 \\
FTPI-MINN-16/20 \\
\end{flushright}

\vspace{0.5cm}
\begin{center}
{\Large \bf Precision gaugino mass measurements as a probe\\[3pt]
of large trilinear soft terms at ILC
}\\ 
\vspace{1.2cm} \renewcommand{\thefootnote}{\fnsymbol{footnote}}
{\large Kyu Jung Bae$^1$\footnote[1]{Email: bae@hep-th.phys.s.u-tokyo.ac.jp},
Howard Baer$^{2,3,4}$\footnote[2]{Email: baer@nhn.ou.edu },
Natsumi Nagata$^{4}$\footnote[3]{Email: nagat006@umn.edu}
and Hasan Serce$^2$\footnote[4]{Email: serce@ou.edu}
}\\ 
\vspace{1.2cm} \renewcommand{\thefootnote}{\arabic{footnote}}
{\it 
$^1$Dept. of Physics, University of Tokyo, Bunkyo-ku, Tokyo 113--0033, Japan \\[3pt]
$^2$Dept. of Physics and Astronomy,
University of Oklahoma, Norman, OK 73019, USA \\[3pt]
}
{\it 
$^3$Kavli Institute for Theoretical Physics, \\ 
University of California, Santa Barbara, CA 93106 USA \\[3pt]
} 
{\it 
$^4$William I. Fine Theoretical Physics Institute, School of Physics and
 Astronomy, \\
University of Minnesota, Minneapolis, MN 55455, USA \\[3pt]
}

\end{center}

\vspace{0.5cm}
\begin{abstract}
\noindent 
In supersymmetric models with radiatively-driven naturalness, 
higgsino-like electroweak-inos (EW-inos) are expected to lie 
in a mass range 100--300 GeV, the lighter the more natural.
Such states can be pair-produced at high rates at ILC where their masses
are nearly equal to the value of the superpotential $\mu$ parameter
while their mass splittings depend on the gaugino masses $M_1$ and $M_2$. 
The gaugino masses in turn depend on trilinear soft terms---the $A$
 parameters, which are expected to lie in the multi-TeV range owing to
 the 125 GeV Higgs mass---via two-loop contributions to renormalization
 group running. We examine the extent to which ILC is sensitive to large
 $A$-terms via precision 
EW-ino mass measurement. Extraction of gaugino masses at the percent
level should allow 
%interesting probes of gaugino mass unification or,
%assuming unified gaugino masses, 
for interesting probes of large trilinear soft SUSY breaking terms
under the assumption of unified gaugino masses.
%{\color{blue} (KJ: We would better focus on A-term effects on gaugino masses. 
%I erased ``interesting probes of gaugino mass unification" part.)}

\vspace*{0.8cm}
%\noindent PACS numbers: 12.60.Jv,14.80.Va,14.80.Ly

\end{abstract}

\end{titlepage}

\section{Introduction}
\label{sec:intro}

The initial spate of results from LHC Run 1 at $\sqrt{s}=7$--8 TeV and Run 2
with $\sqrt{s}=13$ TeV have been delivered and have caused a paradigm shift
in expected phenomenology of supersymmetric (SUSY) models. In pre-LHC
years, a rather light spectrum of sparticle masses was generally
expected on the basis of naturalness: that weak-scale SUSY should not
lie too far beyond the weak scale as typified by the $W,\ Z$ (and
ultimately $h$) masses, where $m_{\rm weak}\simeq m(W,Z,h)\simeq 100$
GeV. These expectations were backed up by calculations of upper bounds
on sparticle masses using the Barbieri--Giudice measure
\cite{BG,DG,AC,ellis,ross}  
\be
\Delta_{\rm BG}\equiv \max_i
\left\vert \frac{\partial \log m_Z^2}{\partial \log p_i}\right\vert ~,
\ee
where $m_Z$ is the $Z$-boson mass and the $p_i$ label fundamental
parameters of the theory, usually taken to be  
the unified soft SUSY breaking terms and the superpotential $\mu$ parameter.
The upper bounds turned out to be typically in the range of a few hundred GeV: 
for instance, in Ref.~\cite{BG}, it was found that $m_{\tg}\lesssim 350$
GeV for $\Delta_{\rm BG}<30$.

Naive hopes for a rapid discovery of SUSY at LHC were dashed by the
reality of data wherein lately the ATLAS and CMS collaborations have
produced bounds on the gluino mass of $m_{\tg}>1.8$~TeV in simplified
models assuming $\tg\to b\bar{b}\widetilde{Z}_1$ decays \cite{ATLASmultibjets,
Khachatryan:2016kdk}. The effect of direct sparticle mass limits was
compounded by the rather high value of the Higgs-boson mass $m_h\simeq 125$~GeV
\cite{Aad:2015zhl} which was discovered. Such a high Higgs mass required
top squarks in the $10$--100~TeV range for small stop mixing and in the
few TeV range for large stop mixing. While heavy stops were allowed by
the $\Delta_{\rm BG}$ measure in the focus-point region \cite{fp},
they were seemingly dis-allowed by large logarithmic contributions to
$m_h$ as quantified by the high-scale large-log measure $\Delta_{\rm
HS}$ \cite{fathiggs,kitnom} which seemed to require {\it three} third
generation squarks with mass $\lesssim 500$~GeV \cite{oldnsusy}. 
Thus, the LHC data cast a pall on overall expectations for SUSY and
indeed led to doubts as to whether weak scale SUSY did actually provide
nature's solution to the gauge hierarchy problem. In addition, lack of
new physics at LHC cast doubt on the motivation for new accelerators
such as the International Linear $e^+e^-$ Collider (ILC) which would
operate at $\sqrt{s}=0.5$--1~TeV\cite{Baer:2013cma,Moortgat-Picka:2015yla}: 
Would there be any prospect for detection of new particles, 
or would the role of such a machine be mainly to
tabulate assorted precision measurements as a Higgs factory?
Prospects for ILC detection of any SUSY particles \cite{jlc,bmt} seemed
dim, much less embarking on a program of precision SUSY particle
measurements to determine underlying high scale Lagrangian
parameters \cite{bpz}. 

An alternative approach was to scrutinize the validity of the
theoretical naturalness calculations. The most conservative approach to
naturalness arises from the weak scale link between the measured 
value of $m_Z$ and the fundamental SUSY Lagrangian parameters \cite{ltr,rns}.
Minimization of the scalar potential in the minimal supersymmetric
Standard Model (MSSM) leads to the well-known relation \cite{wss}
\bea
\frac{m_Z^2}{2}&=&\frac{m_{H_d}^2+\Sigma_d^d-(m_{H_u}^2+\Sigma_u^u)\tan^2\beta}{\tan^2\beta -1}-\mu^2\label{eq:mzs1} \\
&\simeq& -m_{H_u}^2-\Sigma_u^u-\mu^2
\label{eq:mzs}
\eea
where $\Sigma_u^u$ and $\Sigma_d^d$ denote the 1-loop
corrections (expressions can be found in the Appendix of
Ref.~\cite{rns}) to the scalar potential, $m_{H_u}^2$ and $m_{H_d}^2$
are the Higgs soft masses at the weak scale, and $\tan \beta \equiv \langle H_u \rangle /
\langle H_d \rangle$ is the ratio of the Higgs VEVs.  
The second line obtains for moderate to large values of $\tan\beta \agt 5$ (as required by
the Higgs mass calculation\cite{mhiggs}).
SUSY models requiring large cancellations between the various terms on the
right-hand-side of Eq.~\eqref{eq:mzs} to reproduce the measured value of
$m_Z^2$ are regarded as unnatural, or fine-tuned. In contrast, 
SUSY models which generate terms on the RHS of Eq.~\eqref{eq:mzs} which are all
less than or comparable to $m_{\rm weak}$ are regarded as natural. Thus, the
{\it electroweak} naturalness measure $\delew$ is defined as \cite{ltr,rns}
\be
\delew\equiv \text{max}|{\rm each\ additive\ term\ on\ RHS\ of\
Eq.}~\eqref{eq:mzs1}|/(m_Z^2/2). 
\ee
Including the various radiative corrections, over 40 terms contribute.
Neglecting radiative corrections, and taking moderate-to-large $\tan\beta \gtrsim 5$, 
then $m_Z^2/2 \sim-m_{H_u}^2-\mu^2$ so the main
criterion for naturalness is that {\it at the weak scale} 
\bi
\item $m_{H_u}^2\sim -m_Z^2$ and 
\item $\mu^2\sim m_Z^2$ \cite{ccn}.
\ei
The value of $m_{H_d}^2$ (where $m_A\sim m_{H_d}(\text{weak})$ with
$m_A$ being the mass of the CP-odd Higgs boson) can lie in the TeV range
since its contribution to the RHS of Eq.~\eqref{eq:mzs} is suppressed by
$1/\tan^2\beta$. The largest radiative corrections typically come from the top squark
sector. Requiring highly mixed TeV-scale top squarks minimizes
$\Sigma_u^u(\tst_{1,2})$ whilst lifting the Higgs mass $m_h$ to $\sim
125$~GeV \cite{rns}. This framework is called the radiatively-driven
natural SUSY (RNS) \cite{ltr,rns} scenario. 

Using $\Delta_{\rm EW}<30$ or better than 3\% fine-tuning\footnote{For
higher values of $\Delta_{\rm EW}$, high fine-tuning sets in and is
displayed visually in Fig.~2 of Ref.~\cite{upper}. } 
then instead of earlier upper bounds, it is found that
\bi
\item $m_{\tg}\alt 3$--4~TeV,
\item $m_{\tst_1}\alt 3$ TeV and
\item $m_{\tw_1,\tz_{1,2}}\alt 300$ GeV.
\ei
Thus, gluinos and squarks may easily lie beyond the current reach of LHC
at little cost to naturalness while only the higgsino-like lighter
charginos and neutralinos are required to lie near the weak scale. The
lightest higgsino $\widetilde{Z}_1$ comprises a portion of the dark
matter and would escape detection at LHC. Owing to their compressed
spectrum with mass gaps $m_{\tw_1}-m_{\tz_1}\sim m_{\tz_2}-m_{\tz_1}\sim
10$--20~GeV, the heavier higgsinos are difficult to see at LHC owing to
the rather small visible energy released from their three body decays
$\tw_1\to f\bar{f}'\tz_1$ and $\tz_2\to f\bar{f}\tz_1$ (where the $f$
stands for SM fermions). 
At the ILC, on the other hand, direct searches for higgsino-like neutralinos and 
charginos are more promising since the ILC background is much simpler.
We will briefly review ILC direct searches for neutralinos and charginos
in Sec.~\ref{sec:ilc}.

The apparent conflict between $\Delta_{\rm EW}$ and $\Delta_{\rm BG}$
was resolved when it was pointed out that $\Delta_{\rm BG}$ was
typically applied to low energy effective SUSY theories wherein the soft
terms were introduced as independent parameters to parametrize the
unknown dynamics of hidden sector SUSY breaking. If instead the soft
terms are all calculable {\it e.g.} as multiples of the gravitino mass
$m_{3/2}$ as in gravity-mediation, then positive and negative high scale
contributions to $m_Z^2$ cancel and the measure reduces to $\Delta_{\rm
EW}$ \cite{comp3,seige,xt}. Likewise, if one combines all dependent
contributions in the large log measure $\Delta_{\rm HS}$ then
cancellations between $m_{H_u}^2(\Lambda)$ and $\delta m_{H_u}^2$ are
possible so that $\Delta_{\rm HS}\simeq \Delta_{\rm EW}$: in this case 
the large negative correction $\delta m_{H_u}^2$ is used to drive the 
large GUT-scale soft term $m_{H_u}^2$ to a natural value at the weak scale.

\begin{figure}[t]
\begin{center}
 \includegraphics[clip, width = 0.7 \textwidth]{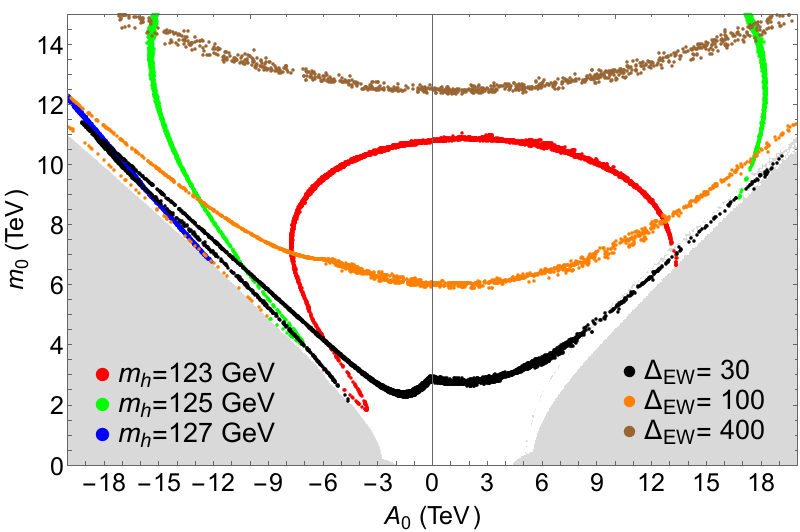}
\caption{The $A_0$ vs. $m_0$ parameter plane for 
$\mu =150$ GeV, $m_{1/2}=1$ TeV, $\tan\beta =10$ and $m_A=2$ TeV.
We show contours of $m_h=123$, 125 and 127 GeV (red, green, blue).
We also show contours of $\Delta_{\rm EW}=30$, 100 and 400. The
 intersection of $m_h\simeq 125$ GeV and low $\Delta_{\rm EW}$ occurs
 along the edges of the allowed region at very large values of $A_0$.
}
\label{fig:plane}
\end{center}
\end{figure}

A grand overview plot of the natural SUSY parameter space is shown in
Fig. \ref{fig:plane} where we show contours of $m_h=123$, 125 and 127
GeV (red, green and blue contours) in the $A_0$ vs. $m_0$ plane for $\mu
=150$ GeV, $m_{1/2}=1$ TeV, $\tan\beta =10$ and $m_A=2$ TeV. We also
show contours of electroweak naturalness $\Delta_{\rm EW}=30$, 100 and
400 (black, orange and brown contours). While it is possible to be highly
natural for small $m_0$ and small $A_0$ (as expected in pre-LHC days),
this region generates a rather light Higgs mass of typically $m_h\sim
115$--120~GeV. However, it is also possible to be highly natural if one
moves to multi-TeV values of $m_0$ and $A_0$ just above the gray
excluded region (where CCB=charge or color breaking minima occur). In
fact, in this large $m_0$ and $A_0$ region, we also see that the Higgs
mass moves up to allowed values $\sim 125$~GeV. Thus, the intersection
of these regions---highly natural with $m_h\sim 125$~GeV---requires
multi-TeV values of $m_0$ and $A_0$. In fact, in a recent paper
\cite{landscape}, it was suggested that in models with $\mu\ll m_{\rm
SUSY}$, the string landscape statistically favors as large as possible 
values of soft terms which are at the same time consistent with the anthropic requirement of 
$m_{\rm weak}\sim 100$~GeV. In this region, the EW symmetry is barely broken
leading to $m_{H_u}^2({\rm weak})\sim -m_Z^2$.
Then very large, multi-TeV values of $m_0$ and $A_0$ are to be expected.

Since $m_0$ sets the squark and slepton mass scale, we expect if we live
in this region then the matter scalars will likely be out of LHC reach. 
Top squarks are also likely to be beyond both LHC and ILC reach  so it
will be difficult to ascertain whether the trilinear terms $A_0$ are
indeed large. 
However, such a large $A_0$ modifies the gaugino masses via 
two loop terms in the renormalization
group evolution.
If the ILC measures gaugino masses at the percent level, 
it is possible to indirectly see the large $A_0$ which is required for natural SUSY.
In this paper, we advocate precision measurements of the
lightest chargino and neutralino masses as a means to test the
requirement of multi-TeV trilinear soft breaking terms.
%
%%%%%%%%%%%%%%%%%%%%% Figure %%%%%%%%%%%%%%%%%%%%%%%%%%%%%%%%%%
%%%%%%%%%%%%%%%%%%%%%%%%%%%%%%%%%%%%%%%%%%%%%%%%%%%%%%%%%%%%%%%%

%%%%%%%%%%%%%%%%%%%%%%%%%%%%%%%%%%%%%%%%%%%%%%%%%%%%%%%%%%%%%%%%
\section{Review of radiatively-driven natural SUSY at ILC}
\label{sec:ilc}
%%%%%%%%%%%%%%%%%%%%%%%%%%%%%%%%%%%%%%%%%%%%%%%%%%%%%%%%%%%%%%%%%%%%

While gluino masses may range up to 4~TeV for $\Delta_{\rm EW}<30$
\cite{rns,upper,guts}, the $5\sigma$ reach of LHC14 for gluino pair production 
with $\sim 1000$~fb$^{-1}$ extends only to about 2~TeV while the 95\% CL
exclusion for $\sim 3000$~fb$^{-1}$ extends to about 2.8~TeV. Other signal
channels such as same sign diboson production\cite{ssdb} from
$pp\to\tw_2\tz_4$ or $pp\to \tz_1\tz_2j\to \ell^+\ell^- j+E^{\rm
miss}_{\rm T}$ \cite{kribs,azar,cheng} should extend the LHC reach
for natural SUSY spectrum\cite{lhc2}. 

Meanwhile, construction of the ILC would be highly propitious for
natural SUSY. If natural SUSY is correct, then ILC would likely be a
{\it higgsino factory} where the reactions $e^+e^-\to \tw_1^+\tw_1^-$
and $\tz_1\tz_2$ would occur at high rates provided
$\sqrt{s}>2m(\text{higgsino})$ and the semi-compressed decays would be
easily visible in the clean environment of $e^+e^-$
collisions.\footnote{Even in the highly compressed  (but less natural)
case of sub-GeV EW-ino mass gaps, precision mass measurements are
possible by using initial state photon radiation \cite{jenny}.}

A detailed phenomenological study of ILC measurements for RNS
has been presented in Ref.~\cite{rnsatilc}. More detailed studies are in
progress using realistic detector simulations and more thorough
background generation \cite{jackie}.
Briefly, the reaction $e^+e^-\to \tw_1^+\tw_1^-$ followed by one
$\tw_1\to \ell\nu_\ell\tz_1$ and the other $\tw_1\to q\bar{q}'\tz_1$
will allow for the $m_{\tw_1}-m_{\tz_1}$ mass gap extraction via
measurement of the dijet invariant mass. In addition, measurement of the
kinematic lower and upper endpoints of the dijet energy $E(jj)$ distribution will
allow for precision determination of $m_{\tw_1}$ and  $m_{\tz_1}$\cite{jlc,bmt,rnsatilc}. 
Typically these endpoint measurements are expected to yield EW-ino masses 
at the percent level at ILC. 
In addition, using the variable beam polarization and the variable beam
energy---for instance to do threshold scans---may allow sub-percent
precision on these mass values\cite{ilctdr}.  
%{\color{blue} (KJ: How realistic is the variable beam experiment? 
%If it is possible, we would better cite some references.)}

Likewise, the reaction $e^+e^-\to \tz_1\tz_2$ followed by
$\tz_2\to\ell^+\ell^-\tz_1$ will allow the $m_{\tz_2}-m_{\tz_1}$ mass
gap to be extracted from the $m(\ell^+\ell^-)$ upper bound. The
endpoints of the $E(\ell^+\ell^- )$ distribution should allow extraction of $m_{\tz_2}$
and $m_{\tz_1}$ to percent or better precision.

While the values of $m_{\tw_1}$, $m_{\tz_2}$ and $m_{\tz_1}$ are all
expected to be $\sim \mu$, the mass gaps will depend on the bino and
wino contamination of the mainly-higgsino eigenstates. Thus, it is
expected that the weak scale bino $M_1$ and wino $M_2$ masses can be
extracted from precision measurements. These virtual mass extractions
will only be aided if measurements of $\tan\beta$ can be gained by
studying deviations in the light Higgs $h$ decay
modes \cite{higgs,Endo:2015oia, us, Kakizaki:2015zva} or
if the heavy Higgs bosons $H$ or $A$ are accessible to LHC (or later
$\sqrt{s}=1$ TeV ILC) searches. In addition, other EW-ino reactions such
as $e^+e^-\to \tz_1\tz_3$ may ultimately be accessible which would
involve direct production of the mainly bino $\tz_3$ state. At even
higher energies, it may be possible to directly produce the mainly wino
state $\tz_4$ via $e^+e^-\to\tz_1\tz_4$. What is likely is that for RNS
with light higgsinos, are very rich program of assorted SUSY
measurements awaits the ILC program. 

\section{Testing multi-TeV trilinears via determination of gaugino masses}
\label{sec:A0}

In this section, we present large $A$-term effects on gaugino masses via the renormalization group evolution from the high energy scale to the weak scale.
The two-loop renormalization group equations (RGEs) for the U(1),
SU(2)$_L$ and SU(3)$_C$ gaugino masses in the $\overline{\rm DR}$ scheme
read \cite{mv}~:
\begin{align}
 \frac{dM_1}{dt}&=\frac{2}{16\pi^2}\frac{33}{5}g_1^2M_1 
+\frac{2g_1^2}{(16\pi^2)^2}\biggl[{199\over 25}g_1^2(2M_1)+
{27\over 5}g_2^2(M_1+M_2)+{88\over 5}g_3^2(M_1+M_3)
\nonumber\\ 
&+  {26\over 5}f_t^2(A_t-M_1)+
{14\over 5}f_b^2(A_b-M_1)+{18\over 5}f_\tau^2(A_\tau-M_1)
%+{6\over 5}f_\nu^2 (A_\nu-M_1)
\biggr] ~,
\label{eq:M1}
\\
\frac{dM_2}{dt}&=\frac{2}{16\pi^2}g_2^2M_2 
+\frac{2g_2^2}{(16\pi^2)^2}\biggl[{9\over 5}g_1^2(M_2+M_1)+
25g_2^2(2M_2)+24g_3^2(M_2+M_3)\nonumber\\ 
&+  6f_t^2(A_t-M_2)+
6f_b^2(A_b-M_2)+2f_\tau^2(A_\tau-M_2)
%+2f_\nu^2 (A_\nu-M_2)
\biggr] ~,
\label{eq:M2}
\\
\frac{dM_3}{dt}&=\frac{2}{16\pi^2}(-3)g_3^2M_3 
+\frac{2g_3^2}{(16\pi^2)^2}\biggl[{11\over 5}g_1^2(M_3+M_1)+
9g_2^2(M_3+M_2)+14g_3^2(2M_3)\nonumber\\ 
&+  4f_t^2(A_t-M_3)+
4f_b^2(A_b-M_3)\biggr]~,
\label{eq:M3}
\end{align}
where $t \equiv \ln Q$ with $Q$ the renormalization scale; $g_1$, $g_2$,
and $g_3$ are the U(1), SU(2)$_L$ and SU(3)$_C$ gauge coupling
constants, respectively (we use the SU(5) normalization for $g_1$);
$f_t$, $f_b$ and $f_\tau$ are the $t$, $b$ and $\tau$ Yukawa couplings,
respectively.\footnote{The third generation neutrino Yukawa coupling
$f_\nu$ also contributes to the running of $M_1$ and $M_2$ above the
right-handed neutrino mass scale since in simple SO(10) based grand
unified theories (GUTs) we expect $f_\nu =f_t$ at the GUT scale. We 
however ignore its effects in the following analysis---if we assume the third
generation neutrino mass to be  $\sim 0.1$~eV, the corresponding
right-handed neutrino mass should be $\sim 3 \times 10^{14}$~GeV. In
this case, the inclusion of the neutrino Yukawa contribution changes
$M_1$ and $M_2$ by $\lesssim 0.05$\%, which is totally negligible in the
present discussion. }
%We keep in addition the contribution from the third
%generation neutrino Yukawa coupling $f_\nu$ since in simple SO(10) based
%grand unified theories (GUTs) they can yield an additional contribution
%where $f_\nu =f_t$ at the GUT scale, $m_{\rm GUT}$. 
While generically
the one-loop contributions to $dM_i/dt$ dominate, in the case where the
soft trilinears are large, then the two loop contributions, which
include the $A_i$ terms, can make a significant effect on the $M_i$
running\cite{bklss}.  

\begin{figure}[t]
\begin{center}
 \includegraphics[clip, width = 0.7 \textwidth]{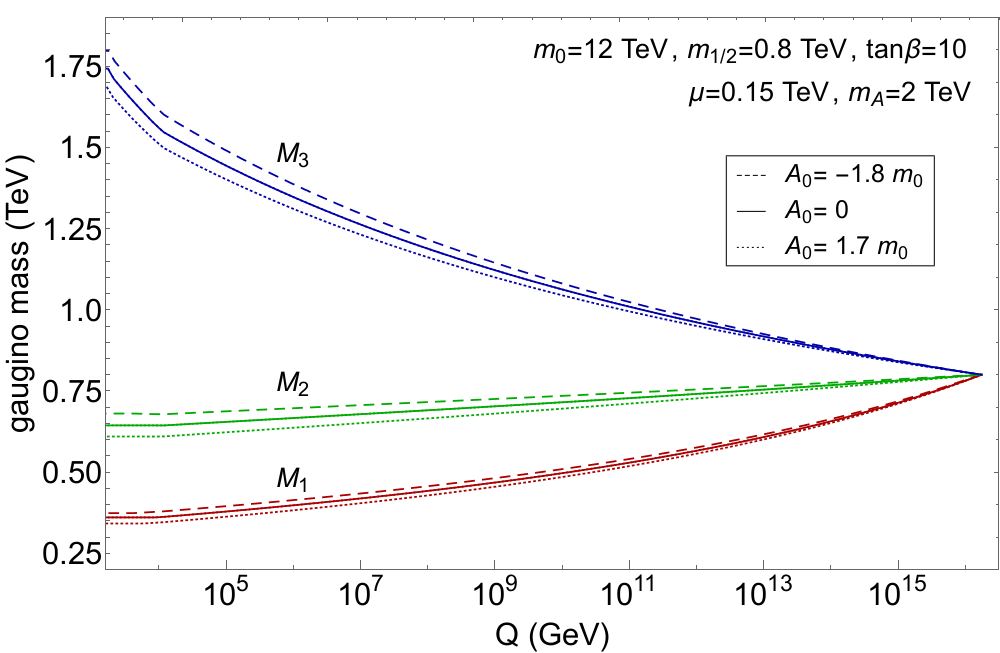}
\caption{Evolution of gaugino masses in radiatively-driven natural SUSY
for large positive and negative and small trilinear soft SUSY breaking terms.
 }
\label{fig:Mi}
\end{center}
\end{figure}

In Fig.~\ref{fig:Mi}, we show the RG trajectories of the gaugino masses
$M_i$ versus energy scale $Q$ starting from $Q=m_{\rm GUT}$ down to the
weak scale. We work in the 2-extra-parameter non-universal Higgs model
(NUHM2) \cite{nuhm2} where matter scalars have unified masses $m_0$ at
$Q=m_{\rm GUT}$ but where Higgs soft terms $m_{H_u}$ and $m_{H_d}$ are
independent since they necessarily live in different GUT multiplets. For
convenience, we trade the GUT scale parameters $m_{H_u}^2$ and
$m_{H_d}^2$ for the weak-scale parameters $\mu$ and $m_A$ so that the
model parameter space is given by 
\be
m_0, m_{1/2}, A_0, \tan\beta, \mu\ {\rm and} \ m_A \ \ ({\rm NUHM2}).
\ee
We use the Isajet 7.85 code for sparticle mass spectrum generation in the NUHM2 model\cite{isajet}.

In Fig.~\ref{fig:Mi}, we adopt parameter choices
$m_0=12$~TeV, $m_{1/2}=0.8$~TeV, $\tan\beta =10$, $\mu=0.15$~TeV and
$m_A=2$~TeV. The solid lines show the $M_i$ ($i=1, 2, 3$) evolution for
$A_0=0$. The dashed lines show the $M_i$ evolution for $A_0=-1.8
m_0$. In this case, the $A$-terms yield a negative contribution to the
right-hand-sides of Eq's~(\ref{eq:M1}--\ref{eq:M3}) thus noticeably
steepening the RG slope of the $M_3$ running and reducing the slopes of
$M_1$ and $M_2$. The effect is especially noticeable for $M_2$ and
$M_3$. Meanwhile, the dotted curves are plotted for $A_0=+1.7 m_0$. In
this case, the two-loop contributions to the $M_i$ running are positive
thus steepening the slopes for $M_1$ and $M_2$ while decreasing the
slope for $M_3$. The extraction of $M_1$ and $M_2$ to high precision, 
%followed by 
%extrapolation via RG running to high energies, 
provided that the gaugino masses are unified at the GUT scale,
%would be an excellent test of
%the hypothesis of gaugino mass unification.
would be an excellent measure of large $A$-terms 
although we do not directly detect scalar quarks or scalar leptons.
%{\color{blue} (KJ: Here again, it seems good to focus on A-term effects on gaugino masses. I slightly modified the sentence.)}
%
%%%%%%%%%%%%%%%%%%%%% Figure %%%%%%%%%%%%%%%%%%%%%%%%%%%%%%%%%%
%%%%%%%%%%%%%%%%%%%%%%%%%%%%%%%%%%%%%%%%%%%%%%%%%%%%%%%%%%%%%%%%

\begin{figure}[t]
\begin{center}
\subfigure[$M_2/M_1$]
{\includegraphics[clip, width = 0.48 \textwidth]{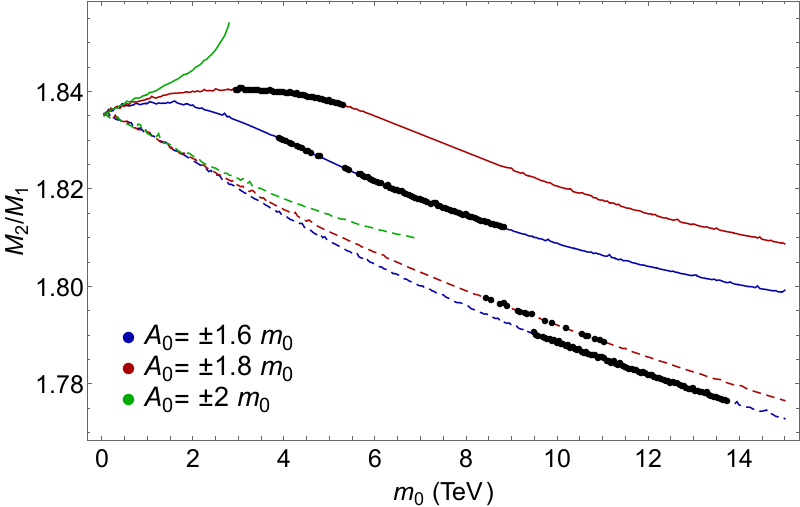}
\label{fig:mA1}}
%\hspace{+0.01\textwidth}
\subfigure[$M_3/M_1$]
{\includegraphics[clip, width = 0.48 \textwidth]{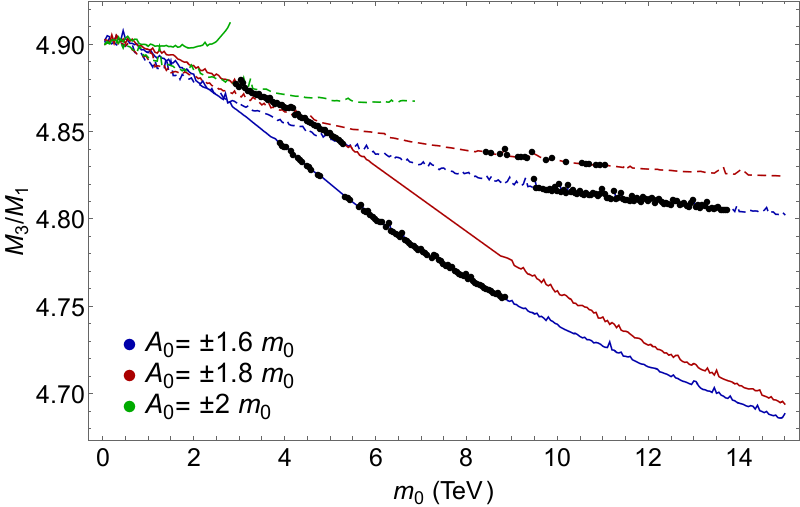}
\label{fig:mA2}}\\
\subfigure[$M_3/M_2$]
{\includegraphics[clip, width = 0.48 \textwidth]{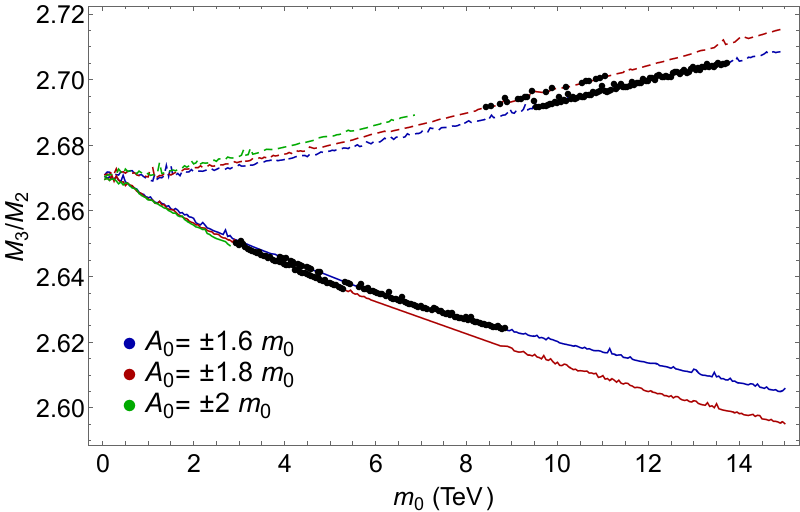}
\label{fig:mA3}}
\caption{Ratios of weak-scale gaugino masses vs. $m_0$ for
$A_0=\pm 1.6 m_0$, $\pm 1.8 m_0$ and $\pm 2m_0$ for $m_{1/2}=1$~TeV, 
$\tan\beta =10$, $\mu=150$~GeV and $m_A=2$~TeV.
 }
\label{fig:ratios_vs_m0}
\end{center}
\end{figure}

In Fig.~\ref{fig:ratios_vs_m0}, we show the ratios of weak-scale gaugino masses
versus $m_0$ which are generated for $m_{1/2}=1$~TeV, $\tan\beta =10$,
$\mu=0.15$~TeV and $m_A=2$~TeV but where $A_0=-1.6 m_0$, $-1.8m_0$ and
$-2m_0$ (solid curves colored blue, red and green respectively) and
$A_0=+1.6 m_0$, $+1.8m_0$ and $+2m_0$ (dashed curves also colored blue,
red and green). The black dotted regions correspond to where $124~{\rm
GeV}<m_h <126$~GeV. The green curves end abruptly when the parameters
conspire to give CCB minima in the scalar potential. Of direct relevance
to ILC is the first frame, Fig.~\ref{fig:mA1}, which shows the ratio
$M_2/M_1$ which can be extracted from precision measurements of EW-ino
masses and mixings. The ratio varies by over $\sim 3\%$ on the range
shown. Given sufficient accuracy in the ILC determination of 
$M_1$ and $M_2$, then it should be possible to determine the {\it sign}
of $A_0$ and perhaps even gain information on the magnitude of $A_0$.
%
%%%%%%%%%%%%%%%%%%%%% Figure %%%%%%%%%%%%%%%%%%%%%%%%%%%%%%%%%%
%%%%%%%%%%%%%%%%%%%%%%%%%%%%%%%%%%%%%%%%%%%%%%%%%%%%%%%%%%%%%%%%

In the event that LHC also discovers gluinos via gluino pair production,
then it is possible the gluino mass can be extracted with order a few
percent accuracy\cite{mgluino,frank} (depending on event rate, 
dominant gluino decay modes and backgrounds). 
If the value of $M_3$ can be extracted from the
gluino pole mass,\footnote{See Ref's~\cite{Yamada:2005ua,Martin:2005ch} 
for the two-loop order calculations and Ref.~\cite{Martin:2006ub} 
for the leading three-loop contributions.} 
%{\color{blue}
%(KJ: I added refs for gluino pole mass calculation. 
%If you know more, please correct it or add more references.)}}
then also the ratios $M_3/M_1$ and $M_3/M_2$ should
become relevant. These ratios are shown versus $m_0$ in frames
Figs.~\ref{fig:mA2} and \ref{fig:mA3}. The ratio $M_3/M_2$ especially
shows a significant disparity in values depending on whether the $A_0$
terms are positive or negative.

\begin{figure}[t]
\begin{center}
\subfigure[$M_2/M_1$]
{\includegraphics[clip, width = 0.48 \textwidth]{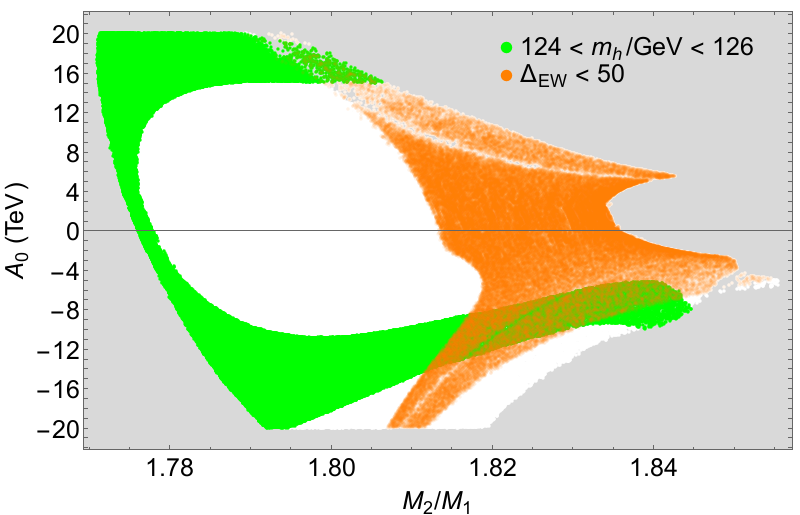}
\label{fig:1}}
\subfigure[$M_3/M_1$]
{\includegraphics[clip, width = 0.48 \textwidth]{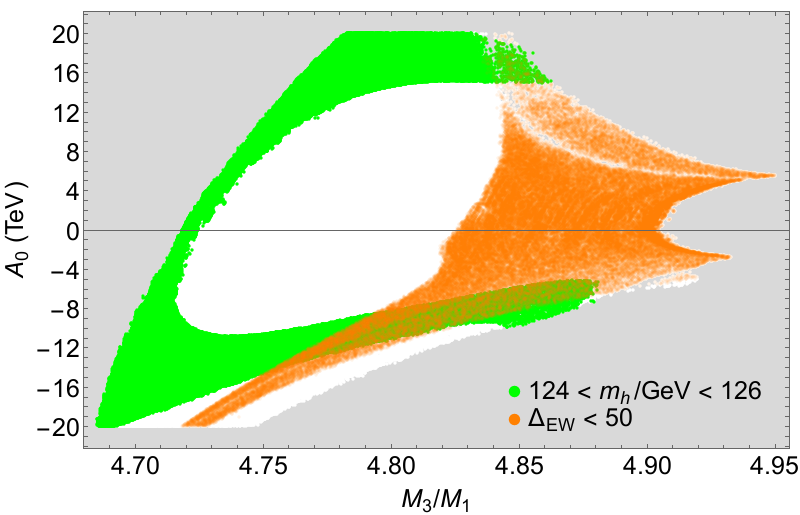}
\label{fig:2}}
\subfigure[$M_3/M_2$]
{\includegraphics[clip, width = 0.48 \textwidth]{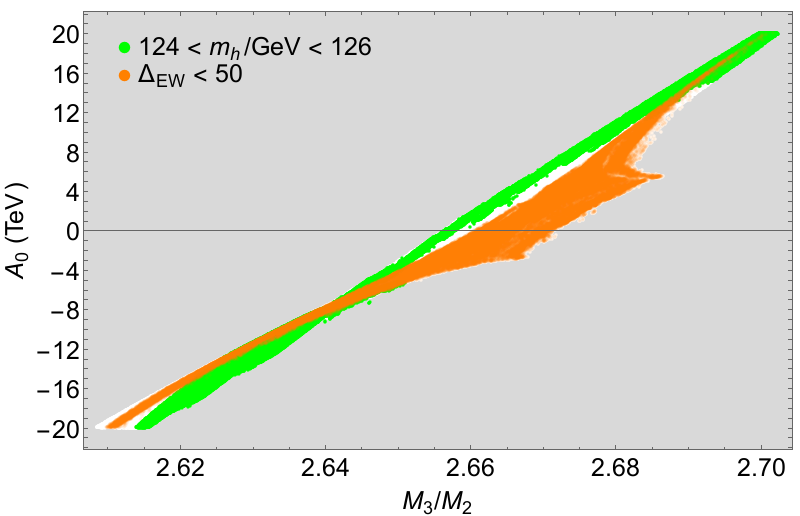}
\label{fig:3}}
\caption{Ratios of weak scale gaugino masses vs. $A_0$ for 
$m_{1/2}=1$ TeV, $\tan\beta =10$, $\mu=150$ GeV and $m_A=2$ TeV
but with $m_0$ scanned over the range 0--15 TeV 
and $-20\ {\rm TeV}<A_0<+20$ TeV.
 }
\label{fig:ratios_vs_A0}
\end{center}
\end{figure}

In Fig's~\ref{fig:ratios_vs_A0} we show the weak-scale gaugino mass
ratios versus variation in $A_0$ for $m_{1/2}=1$~TeV, $\tan\beta =10$,
$\mu=0.15$~TeV and $m_A=2$~TeV but where $m_0$ is scanned over the range
$m_0: 0-15$~TeV and $-20~\text{TeV}<A_0<+20$~TeV. The gray regions are
disallowed by either CCB scalar potential minima or no EWSB. The green
regions yield a light Higgs mass with $124~\text{GeV}<m_h<126$~GeV while
the orange regions indicate $\Delta_{\rm EW}<50$. For Fig.~\ref{fig:1},
we see two intersecting regions: one for positive $A_0$ and one for
negative $A_0$. The low fine-tuned regions which are consistent with the
measured value of $m_h$ occur at large $|A_0|$ values indicating
large mixing in the stop sector. The ratio $M_2/M_1 \agt 1.81$ for the
case of $A_0$ large negative while $M_2/M_1\alt 1.81$ for $A_0$ large
positive. We also notice that these two regions can  be sufficiently
separated from those which predict the correct Higgs mass but a large
value of $\delew$. 
In particular, $A_0\simeq 0$ points with $m_h\simeq 125$ GeV 
predict $M_2/M_1 \lesssim 1.78$, 
which can be distinguished from the above two regions if
$M_2/M_1$ is measured with $\sim 1$\% accuracy. 
Thus, assuming unified gaugino masses, it appears precision
measurements of EW-ino masses and mixings will be sensitive to details of
large trilinear soft terms. 
%
%%%%%%%%%%%%%%%%%%%%% Figure %%%%%%%%%%%%%%%%%%%%%%%%%%%%%%%%%%
%%%%%%%%%%%%%%%%%%%%%%%%%%%%%%%%%%%%%%%%%%%%%%%%%%%%%%%%%%%%%%%%

In Fig.~\ref{fig:2}, we show the ratio $M_3/M_1$ for the case where LHC14 
can discover the gluino and gain an estimate of $M_3$. Here, we see the
intersecting orange/green regions occur for $M_3/M_1\agt 4.84$ for $A_0$
large positive while $M_3/M_1\sim 4.75$--$4.88$ for $A_0$ large
negative. While some range of $M_3/M_1$ overlaps between these two
cases, only the large negative $A_0$ accesses the smaller range
$M_3/M_1\alt 4.85$. 
In Fig.~\ref{fig:3}, we plot the ratio $M_3/M_2$
versus $A_0$. Here we see that large positive values of $A_0$ yield
$M_3/M_2\sim 2.68$--$2.7$ while large negative $A_0$ yields instead
$M_3/M_2\sim 2.62$--$2.65$. The two cases appear distinguishable to
combined ILC and LHC precision measurements.

Before concluding this section, we discuss some possible uncertainties 
in our calculations. In the calculation shown above, we assume the gaugino
mass unification at the GUT scale. However, this can be spoiled by
GUT-scale threshold corrections and by Planck-scale suppressed higher
dimensional operators \cite{Hisano:1993zu, Tobe:2003yj}. These effects
are expected to be $\lesssim 1$\%; however, they can be significant if the
$A$- or $B$-terms for the GUT Higgs fields are much larger than gaugino
masses, and/or if there are large representations of the GUT gauge
group.\footnote{We however note that if top squarks are discovered in
future collider experiments, we may infer $m_0$ and $A_0$ from the
measurements of their masses, and in this case precision gaugino mass
measurements in turn allow us to extract the soft parameters for the GUT Higgs
fields via the GUT threshold corrections, just like the precise
determination of the gauge couplings enables us to extract the mass spectrum
of the GUT-scale fields via the GUT threshold corrections to the
gauge couplings \cite{Hisano:1992mh}. In this sense, precision gaugino
mass measurements are of importance even if the $A$-terms are directly
measured.  } 

\begin{figure}[t]
\begin{center}
\subfigure[gauge coupling running]{
\includegraphics[clip, width = 0.48 \textwidth]{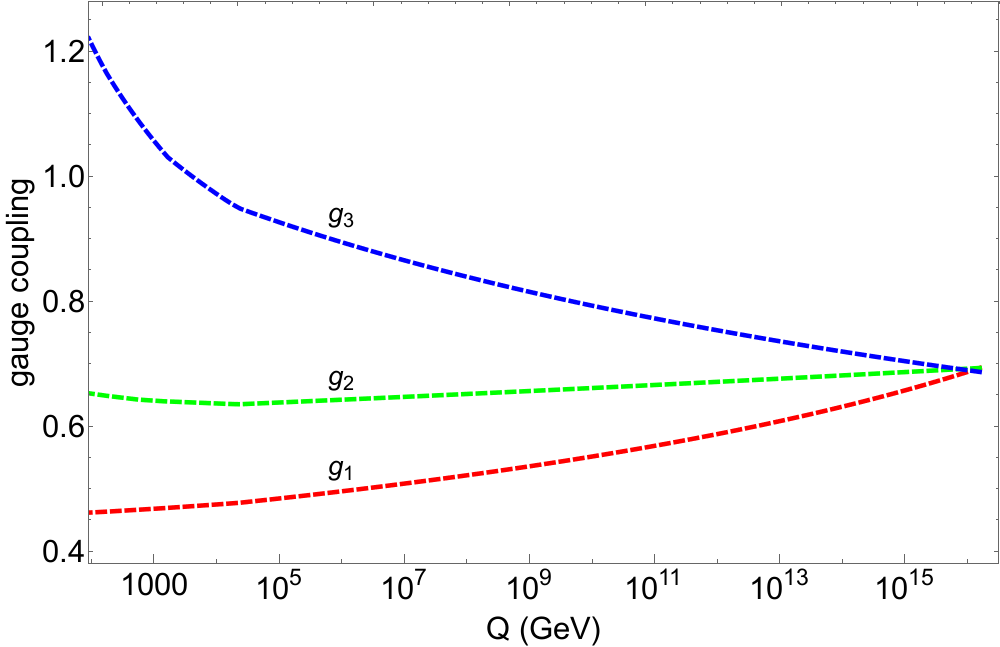}}
\subfigure[gaugino mass running]{
\includegraphics[clip, width = 0.48 \textwidth]{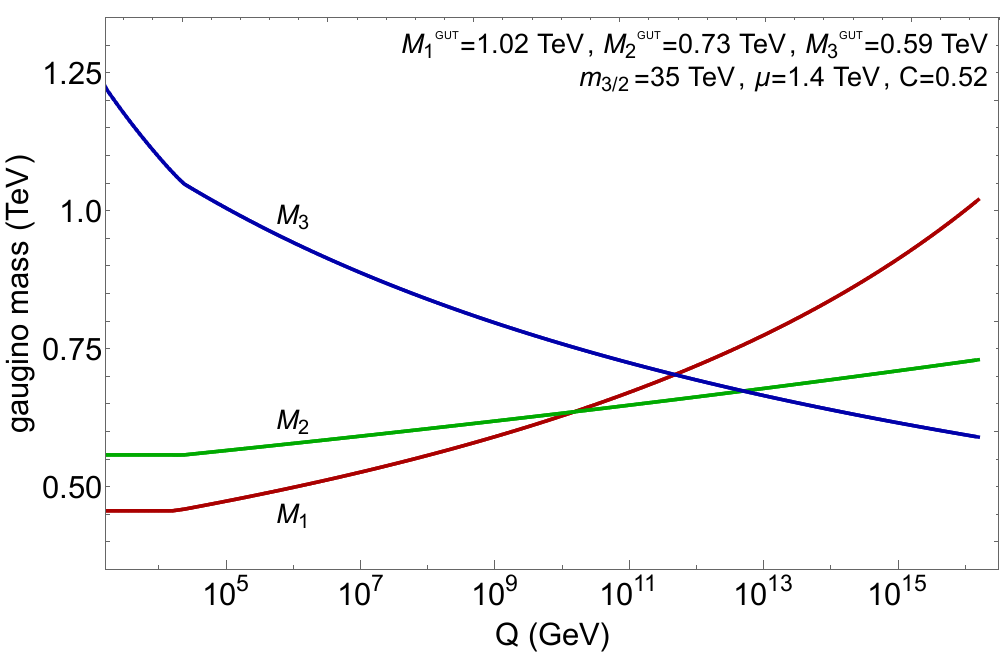}}
\caption{In ({a}), we show the running of gauge couplings and
in ({b}) we show running of gaugino masses in the 
G2MSSM model of Ref.~\cite{g2mssm} with $m_0=24.2$ TeV, $A_0=25.2$ TeV (or $m_{3/2}=35$ TeV, $C=0.52$),
$\tan\beta =8$, $\mu=1.4$ TeV and $M_1(m_{\rm GUT})=1.02$ TeV,
 $M_2(m_{\rm GUT})=0.73$ TeV and $M_3(m_{\rm GUT})=0.59$ TeV.
\label{fig:g2mssm}
 }
\end{center}
\end{figure}

In the event that gaugino masses are not unified, then
extrapolation of the measured values of $M_1$ and $M_2$ to high scales will
intersect at some point other than $Q=m_{\rm GUT}$ where the gauge couplings unify.
This situation is illustrated in Fig.~\ref{fig:g2mssm} where in 
frame ({a}) we show the running of gauge couplings while in frame 
({b}) we show the running of gaugino masses from the 
compactified $M$-theory model of Ref.~\cite{g2mssm}.
If such a theory were discovered, then the precision measurements
of $M_1$ and $M_2$ at LHC and/or ILC would be extrapolated in energy 
to meet at a unification point other than the energy scale 
where the gauge couplings unify.
If $M_3$ is also measured, then the non-unification of all three
gaugino masses would be apparent.
%%%%%%%%%%%%%%%%%%%%% Figure %%%%%%%%%%%%%%%%%%%%%%%%%%%%%%%%%%
%%%%%%%%%%%%%%%%%%%%%%%%%%%%%%%%%%%%%%%%%%%%%%%%%%%%%%%%%%%%%%%%

%\TODO{If we also look into the deviations in the gaugino couplings, we
%may know if sfermion mass scale is very high or not, without suffering
%from the effects from the GUT-scale physics. Let us discuss it later. }

Precision gaugino mass measurements also play an important role in
testing the split-SUSY type mass spectrum \cite{Wells:2003tf}, where
soft masses are taken to be ${\cal O}(100$--1000)~TeV and gaugino masses
and $A$-terms are
suppressed by loop factors, with which the 125~GeV Higgs mass can be
obtained \cite{Giudice:2011cg}. In this case, gaugino masses are induced
via anomaly mediation and proportional to the corresponding gauge
coupling beta functions \cite{Randall:1998uk}, which results in
different gaugino mass ratios from those in the unified gaugino mass
case. Deviations from the anomaly-mediation relation are caused by
renormalization group effects below the soft mass scale and threshold
corrections by higgsino--Higgs one-loop diagrams, which enable us to extract
information on the SUSY scale and higgsino/heavy Higgs mass spectrum
through gaugino mass measurements.

%%%%%%%%%%%%%%%%%%%%%%%%%%%%%%
%\section{Gaugino couplings}
%%%%%%%%%%%%%%%%%%%%%%%%%%%%%%

%\TODO{To be discussed.}

\section{Conclusions}
\label{sec:conclude}

In the post LHC8 world with an improved understanding of 
electroweak naturalness in SUSY models, then we are directed to expect
the existence of rather light higgsino-like $\tw_1^\pm$ and $\tz_{1,2}$
with mass $\sim 100-250$ GeV, the closer to $m_h$ the better.
Such light EW-inos are difficult to see at LHC 
(due to only soft tracks emerging from their decays) 
but which should be easily visible in the clean environment
of an $e^+e^-$ collider such as ILC with $\sqrt{s}\agt 2m({\rm higgsino})$.
While it is expected that $m_{\tw_1,\tz_{1,2}}\sim |\mu|$, the mass splittings 
$m_{\tw_1}-m_{\tz_1}$ and $m_{\tz_2}-m_{\tz_1}$ are sensitive to the gaugino
masses $M_1$ and $M_2$ via mixing effects.

From the rather high value of $m_h$, we also expect large trilinear
soft terms $A_t$ which contribute to large mixing in the stop sector
and an uplifting of $m_h$. One way to test the presence of 
large trilinear soft terms is to 
look for their influence  on the gaugino mass running 
which occurs at two-loop level. We point out that the sign of $A_0$ 
and its magnitude influence the expected ratios of gaugino masses
which may be measured at ILC. Thus, this work motivates a dedicated program of
precision measurements of EW-ino properties at a machine such as ILC 
which may test Lagrangian parameters well-removed from just those that 
directly determine the EW-ino masses.

\section*{Acknowledgments}

This work was supported in part by the US Department of Energy, Office of High
Energy Physics and in part by the National Science Foundation
under Grant No. NSF PHY11-25915. 
KJB is supported by Grant-in-Aid for Scientific research No. 26104009.
The work of NN is supported by the DOE grant DE-SC0011842 at
the University of Minnesota. HB would like to thank the William I. Fine
Institute for Theoretical Physics (FTPI) at the University of Minnesota
for hospitality while this work was initiated and KITP Santa Barbara
where this work was completed. 
The computing for this project was performed at the 
OU Supercomputing Center for Education \& Research (OSCER) at the 
University of Oklahoma (OU).
%
%%%%%%%%%%%%%%%%%%%%%%%%%%%%%%%%%%%%%%%%%%%%%%%%%%%%%%

%

\begin{thebibliography}{99}
%%%%%%%%%%%%%%%%%%%%%%%%%%%%%%%%%%%%%%%%%%%%%%%%%%%%%%
\small
%***
\bibitem{BG} R.~Barbieri and G.~F.~Giudice,
  %``Upper Bounds on Supersymmetric Particle Masses,''
  Nucl.\ Phys.\ B {\bf 306} (1988) 63.
%
\bibitem{DG} S.~Dimopoulos and G.~F.~Giudice,
  %``Naturalness constraints in supersymmetric theories with nonuniversal soft terms,''
  Phys.\ Lett.\ B {\bf 357} (1995) 573.
%
\bibitem{AC} G.~W.~Anderson and D.~J.~Castano,
  %``Measures of fine tuning,''
  Phys.\ Lett.\ B {\bf 347} (1995) 300;
G.~W.~Anderson and D.~J.~Castano,
  %``Naturalness and superpartner masses or when to give up on weak scale supersymmetry,''
  Phys.\ Rev.\ D {\bf 52} (1995) 1693;
G.~W.~Anderson and D.~J.~Castano,
  %``Challenging weak scale supersymmetry at colliders,''
  Phys.\ Rev.\ D {\bf 53} (1996) 2403.
%
\bibitem{ellis}P.~H.~Chankowski, J.~R.~Ellis and S.~Pokorski,
  %``The Fine tuning price of LEP,''
  Phys.\ Lett.\ B {\bf 423} (1998) 327;
 P.~H.~Chankowski, J.~R.~Ellis, M.~Olechowski and S.~Pokorski,
  %``Haggling over the fine tuning price of LEP,''
  Nucl.\ Phys.\ B {\bf 544} (1999) 39.
%
\bibitem{ross} S.~Cassel, D.~M.~Ghilencea and G.~G.~Ross,
  %``Testing SUSY,''
  Phys.\ Lett.\ B {\bf 687} (2010) 214;
S.~Cassel, D.~M.~Ghilencea, S.~Kraml, A.~Lessa and G.~G.~Ross,
  %``Fine-tuning implications for complementary dark matter and LHC SUSY searches,''
  JHEP {\bf 1105} (2011) 120.

%\cite{ATLASmultibjets}
\bibitem{ATLASmultibjets} 
  The ATLAS collaboration,
  %``Search for pair-production of gluinos decaying via stop and sbottom in events with $b$-jets and large missing transverse momentum in $\sqrt{s}=13$ TeV $pp$ collisions with the ATLAS detector,''
  ATLAS-CONF-2015-067.
  %%CITATION = ATLAS-CONF-2015-067;%%
  %8 citations counted in INSPIRE as of 11 Apr 2016

%\cite{Khachatryan:2016kdk}
\bibitem{Khachatryan:2016kdk} 
  V.~Khachatryan {\it et al.} [CMS Collaboration],
  %``Search for supersymmetry in the multijet and missing transverse momentum final state in pp collisions at 13 TeV,''
  %Submitted to: Phys.Lett.B
  [arXiv:1602.06581 [hep-ex]].
  %%CITATION = ARXIV:1602.06581;%%
  %5 citations counted in INSPIRE as of 11 Apr 2016

%\cite{Aad:2015zhl}
\bibitem{Aad:2015zhl} 
  G.~Aad {\it et al.} [ATLAS and CMS Collaborations],
  %``Combined Measurement of the Higgs Boson Mass in $pp$ Collisions at $\sqrt{s}=7$ and 8 TeV with the ATLAS and CMS Experiments,''
  Phys.\ Rev.\ Lett.\  {\bf 114}, 191803 (2015).
%  doi:10.1103/PhysRevLett.114.191803
%  [arXiv:1503.07589 [hep-ex]].
  %%CITATION = doi:10.1103/PhysRevLett.114.191803;%%
  %377 citations counted in INSPIRE as of 11 Apr 2016

%
\bibitem{fp} J.~L.~Feng, K.~T.~Matchev and T.~Moroi,
  %``Focus points and naturalness in supersymmetry,''
  Phys.\ Rev.\ D {\bf 61} (2000) 075005;
J.~L.~Feng, K.~T.~Matchev and T.~Moroi,
  %``Naturalness reexamined: Implications for supersymmetry searches,''
  hep-ph/0003138;
J.~L.~Feng and D.~Sanford,
  %``A Natural 125 GeV Higgs Boson in the MSSM from Focus Point Supersymmetry with A-Terms,''
  Phys.\ Rev.\ D {\bf 86} (2012) 055015.
%
\bibitem{fathiggs}   R.~Harnik, G.~D.~Kribs, D.~T.~Larson and H.~Murayama,
  %``The Minimal supersymmetric fat Higgs model,''
  Phys.\ Rev.\ D {\bf 70} (2004) 015002.
%
\bibitem{kitnom}   R.~Kitano and Y.~Nomura,
  %``Supersymmetry, naturalness, and signatures at the LHC,''
  Phys.\ Rev.\ D {\bf 73}, 095004 (2006).
%
\bibitem{oldnsusy}  M.~Papucci, J.~T.~Ruderman and A.~Weiler,
  %``Natural SUSY Endures,''
  JHEP {\bf 1209} (2012) 035.
%
\bibitem{Baer:2013cma}
  H.~Baer {\it et al.},
  %``The International Linear Collider Technical Design Report - Volume 2: Physics,''
  arXiv:1306.6352 [hep-ph].
%
\bibitem{Moortgat-Picka:2015yla}
  G.~Moortgat-Pick {\it et al.},
  %``Physics at the e+ e- Linear Collider,''
  Eur.\ Phys.\ J.\ C {\bf 75} (2015) no.8,  371.
%
\bibitem{jlc} T.~Tsukamoto, K.~Fujii, H.~Murayama, M.~Yamaguchi and Y.~Okada,
  %``Precision study of supersymmetry at future linear e+ e- colliders,''
  Phys.\ Rev.\ D {\bf 51} (1995) 3153.
%
\bibitem{bmt} H.~Baer, R.~B.~Munroe and X.~Tata,
  %``Supersymmetry studies at future linear e+ e- colliders,''
  Phys.\ Rev.\ D {\bf 54} (1996) 6735.
%
\bibitem{bpz} G.~A.~Blair, W.~Porod and P.~M.~Zerwas,
  %``Reconstructing supersymmetric theories at high-energy scales,''
  Phys.\ Rev.\ D {\bf 63} (2001) 017703;
G.~A.~Blair, W.~Porod and P.~M.~Zerwas,
  %``The Reconstruction of supersymmetric theories at high-energy scales,''
  Eur.\ Phys.\ J.\ C {\bf 27} (2003) 263.
%
\bibitem{ltr} H.~Baer, V.~Barger, P.~Huang, A.~Mustafayev and X.~Tata,
  %``Radiative natural SUSY with a 125 GeV Higgs boson,''
  Phys.\ Rev.\ Lett.\  {\bf 109} (2012) 161802.
%
\bibitem{rns} H.~Baer, V.~Barger, P.~Huang, D.~Mickelson, A.~Mustafayev and X.~Tata,
  %``Radiative natural supersymmetry: Reconciling electroweak fine-tuning and the Higgs boson mass,''
  Phys.\ Rev.\ D {\bf 87} (2013) 115028.
%
\bibitem{wss} H.~Baer and X.~Tata, {\it Weak Scale Supersymmetry: From
Superfields to Scattering Events},
(Cambridge University Press, 2006).
%
\bibitem{mhiggs} M.~Carena and H.~E.~Haber,
  %``Higgs boson theory and phenomenology,''
  Prog.\ Part.\ Nucl.\ Phys.\  {\bf 50} (2003) 63.
%
\bibitem{ccn} K.~L.~Chan, U.~Chattopadhyay and P.~Nath,
  %``Naturalness, weak scale supersymmetry and the prospect for the observation of supersymmetry at the Tevatron and at the CERN LHC,''
  Phys.\ Rev.\ D {\bf 58} (1998) 096004;
%  [hep-ph/9710473];
S.~Akula, M.~Liu, P.~Nath and G.~Peim,
  %``Naturalness, Supersymmetry and Implications for LHC and Dark Matter,''
  Phys.\ Lett.\ B {\bf 709} (2012) 192;
M.~Liu and P.~Nath,
  %``Higgs boson mass, proton decay, naturalness, and constraints of the LHC and Planck data,''
  Phys.\ Rev.\ D {\bf 87} (2013) 9,  095012.
%
\bibitem{upper} H.~Baer, V.~Barger and M.~Savoy,
  %``Upper bounds on sparticle masses from naturalness or how to disprove weak scale supersymmetry,''
  Phys.\ Rev.\ D {\bf 93} (2016),  035016.
%
\bibitem{comp3} H.~Baer, V.~Barger and D.~Mickelson,
  %``How conventional measures overestimate electroweak fine-tuning in supersymmetric theory,''
  Phys.\ Rev.\ D {\bf 88} (2013) 095013.
%
\bibitem{seige} H.~Baer, V.~Barger, D.~Mickelson and M.~Padeffke-Kirkland,
  %``SUSY models under siege: LHC constraints and electroweak fine-tuning,''
  Phys.\ Rev.\ D {\bf 89} (2014) 115019.
%
\bibitem{xt} A.~Mustafayev and X.~Tata,
  %``Supersymmetry, Naturalness, and Light Higgsinos,''
  Indian J.\ Phys.\  {\bf 88} (2014) 991;
X.~Tata,
  %``Supersymmetry: Aspirations and Prospects,''
  Phys.\ Scripta {\bf 90} (2015) 108001.
%
\bibitem{landscape}H.~Baer, V.~Barger, M.~Savoy and H.~Serce,
  %``The Higgs mass and natural supersymmetric spectrum from the landscape,''
  arXiv:1602.07697 [hep-ph]. 
%
\bibitem{guts} H.~Baer, V.~Barger and M.~Savoy,
  %``Generalized focus point and mass spectra comparison of highly natural SUGRA GUT models,''
  Phys.\ Rev.\ D {\bf 93} (2016),  075001.
%
\bibitem{ssdb} H.~Baer, V.~Barger, P.~Huang, D.~Mickelson, A.~Mustafayev, W.~Sreethawong and X.~Tata,
  %``Same sign diboson signature from supersymmetry models with light higgsinos at the LHC,''
  Phys.\ Rev.\ Lett.\  {\bf 110} (2013),  151801;
H.~Baer, V.~Barger, P.~Huang, D.~Mickelson, A.~Mustafayev, W.~Sreethawong and X.~Tata,
  %``Radiatively-driven natural supersymmetry at the LHC,''
  JHEP {\bf 1312} (2013) 013.
%
\bibitem{kribs} Z.~Han, G.~D.~Kribs, A.~Martin and A.~Menon,
  %``Hunting quasidegenerate Higgsinos,''
  Phys.\ Rev.\ D {\bf 89} (2014),  075007.
%
\bibitem{azar} H.~Baer, A.~Mustafayev and X.~Tata,
  %``Monojet plus soft dilepton signal from light higgsino pair production at LHC14,''
  Phys.\ Rev.\ D {\bf 90} (2014),  115007.
%
\bibitem{cheng} C.~Han, D.~Kim, S.~Munir and M.~Park,
  %``Accessing the core of naturalness, nearly degenerate higgsinos, at the LHC,''
  JHEP {\bf 1504} (2015) 132. 
%
\bibitem{lhc2} H.~Baer, V.~Barger, M.~Savoy and X.~Tata,
  %``Multi-channel assault on natural supersymmetry at the high luminosity LHC,''
  arXiv:1604.07438 [hep-ph].
%
\bibitem{jenny}  M.~Berggren, F.~Brümmer, J.~List, G.~Moortgat-Pick, T.~Robens, K.~Rolbiecki and H.~Sert,
  %``Tackling light higgsinos at the ILC,''
  Eur.\ Phys.\ J.\ C {\bf 73} (2013),  2660.
%
\bibitem{rnsatilc} H.~Baer, V.~Barger, D.~Mickelson, A.~Mustafayev and X.~Tata,
  %``Physics at a Higgsino Factory,''
  JHEP {\bf 1406} (2014) 172.
%
\bibitem{ilctdr} H.~Baer {\it et al.},
  %``The International Linear Collider Technical Design Report - Volume 2: Physics,''
  arXiv:1306.6352 [hep-ph].
%
\bibitem{jackie} J. Yan, S.-L. Lehtinen, J. List, M. Bergren, K. Fujii, T. Tanabe and H. Baer, 
in preparation.
%
\bibitem{higgs} T.~Han, Z.~Liu and J.~Sayre,
  %``Potential Precision on Higgs Couplings and Total Width at the ILC,''
  Phys.\ Rev.\ D {\bf 89} (2014) 113006;
 M.~E.~Peskin,
  %``Estimation of LHC and ILC Capabilities for Precision Higgs Boson Coupling Measurements,''
  arXiv:1312.4974 [hep-ph];  S.~Dawson, A.~Gritsan, H.~Logan, J.~Qian, C.~Tully, R.~Van Kooten, A.~Ajaib and A.~Anastassov {\it et al.},
  %``Working Group Report: Higgs Boson,''
  arXiv:1310.8361 [hep-ex].
%\cite{Endo:2015oia}
\bibitem{Endo:2015oia} 
  M.~Endo, T.~Moroi and M.~M.~Nojiri,
  %``Footprints of Supersymmetry on Higgs Decay,''
  JHEP {\bf 1504}, 176 (2015).
%  doi:10.1007/JHEP04(2015)176
%  [arXiv:1502.03959 [hep-ph]].
  %%CITATION = doi:10.1007/JHEP04(2015)176;%%
  %7 citations counted in INSPIRE as of 13 Apr 2016
%
\bibitem{us} K.~J.~Bae, H.~Baer, N.~Nagata and H.~Serce,
  %``Prospects for Higgs coupling measurements in SUSY with radiatively-driven naturalness,''
  Phys.\ Rev.\ D {\bf 92} (2015),  035006.

%\cite{Kakizaki:2015zva}
\bibitem{Kakizaki:2015zva} 
  M.~Kakizaki, S.~Kanemura, M.~Kikuchi, T.~Matsui and H.~Yokoya,
  %``Indirect reach of heavy MSSM Higgs bosons by precision measurements at future lepton colliders,''
  Int.\ J.\ Mod.\ Phys.\ A {\bf 30}, no. 33, 1550192 (2015).
%  doi:10.1142/S0217751X15501924
%  [arXiv:1505.03761 [hep-ph]].
  %%CITATION = doi:10.1142/S0217751X15501924;%%
  %3 citations counted in INSPIRE as of 13 Apr 2016
%
\bibitem{mv} 
%\cite{Martin:1993yx}
%\bibitem{Martin:1993yx} 
  S.~P.~Martin and M.~T.~Vaughn,
  %``Regularization dependence of running couplings in softly broken supersymmetry,''
  Phys.\ Lett.\ B {\bf 318}, 331 (1993);
%  doi:10.1016/0370-2693(93)90136-6
%  [hep-ph/9308222].
  %%CITATION = doi:10.1016/0370-2693(93)90136-6;%%
  %249 citations counted in INSPIRE as of 12 Apr 2016
%\cite{Yamada:1993ga}
%\bibitem{Yamada:1993ga} 
  Y.~Yamada,
  %``Two loop renormalization of gaugino masses in general supersymmetric gauge models,''
  Phys.\ Rev.\ Lett.\  {\bf 72}, 25 (1994);
%  doi:10.1103/PhysRevLett.72.25
%  [hep-ph/9308304].
  %%CITATION = doi:10.1103/PhysRevLett.72.25;%%
  %51 citations counted in INSPIRE as of 12 Apr 2016
S.~P.~Martin and M.~T.~Vaughn,
  %``Two loop renormalization group equations for soft supersymmetry breaking couplings,''
  Phys.\ Rev.\ D {\bf 50} (1994) 2282.
%
\bibitem{bklss} H.~Baer, S.~Kraml, A.~Lessa, S.~Sekmen and H.~Summy,
  %``Beyond the Higgs boson at the Tevatron: Detecting gluinos from Yukawa-unified SUSY,''
  Phys.\ Lett.\ B {\bf 685} (2010) 72.
%
\bibitem{nuhm2}  D.~Matalliotakis and H.~P.~Nilles,
%``Implications of nonuniversality of soft terms in supersymmetric grand unified theories,''                                                                 
  Nucl.\ Phys.\ B {\bf 435} (1995) 115;
P.~Nath and R.~L.~Arnowitt,
  %``Nonuniversal soft SUSY breaking and dark matter,''                         
  Phys.\ Rev.\ D {\bf 56} (1997) 2820;
J. Ellis, K. Olive and Y. Santoso, Phys. Lett. {\bf B539} (2002) 107;
J. Ellis, T. Falk, K. Olive and Y. Santoso,
Nucl. Phys. {\bf B652} (2003) 259;
H.~Baer, A.~Mustafayev, S.~Profumo, A.~Belyaev and X. Tata,
JHEP{\bf 0507} (2005) 065.
%
\bibitem{isajet} ISAJET, by H.~Baer, F.~Paige, S.~Protopopescu and
X.~Tata, \hepph{0312045}.
%
\bibitem{mgluino} H.~Baer, V.~Barger, G.~Shaughnessy, H.~Summy and L.~t.~Wang,
  %``Precision gluino mass at the LHC in SUSY models with decoupled scalars,''
  Phys.\ Rev.\ D {\bf 75} (2007) 095010.
%
\bibitem{frank} I.~Hinchliffe, F.~E.~Paige, M.~D.~Shapiro, J.~Soderqvist and W.~Yao,
  %``Precision SUSY measurements at CERN LHC,''
  Phys.\ Rev.\ D {\bf 55} (1997) 5520.
%

%\cite{Yamada:2005ua}
\bibitem{Yamada:2005ua} 
  Y.~Yamada,
  %``Two-loop SUSY QCD correction to the gluino pole mass,''
  Phys.\ Lett.\ B {\bf 623}, 104 (2005).
%  doi:10.1016/j.physletb.2005.08.004
%  [hep-ph/0506262].
  %%CITATION = doi:10.1016/j.physletb.2005.08.004;%%
  %15 citations counted in INSPIRE as of 16 May 2016

%\cite{Martin:2005ch}
\bibitem{Martin:2005ch} 
  S.~P.~Martin,
  %``Fermion self-energies and pole masses at two-loop order in a general renormalizable theory with massless gauge bosons,''
  Phys.\ Rev.\ D {\bf 72}, 096008 (2005).
%  doi:10.1103/PhysRevD.72.096008
%  [hep-ph/0509115].
  %%CITATION = doi:10.1103/PhysRevD.72.096008;%%
  %36 citations counted in INSPIRE as of 16 May 2016

%\cite{Martin:2006ub}
\bibitem{Martin:2006ub} 
  S.~P.~Martin,
  %``Refined gluino and squark pole masses beyond leading order,''
  Phys.\ Rev.\ D {\bf 74}, 075009 (2006).
%  doi:10.1103/PhysRevD.74.075009
 % [hep-ph/0608026].
  %%CITATION = doi:10.1103/PhysRevD.74.075009;%%
  %11 citations counted in INSPIRE as of 16 May 2016




%\cite{Hisano:1993zu}
\bibitem{Hisano:1993zu} 
  J.~Hisano, H.~Murayama and T.~Goto,
  %``Threshold correction on gaugino masses at grand unification scale,''
  Phys.\ Rev.\ D {\bf 49}, 1446 (1994).
%  doi:10.1103/PhysRevD.49.1446
  %%CITATION = doi:10.1103/PhysRevD.49.1446;%%
  %36 citations counted in INSPIRE as of 26 Apr 2016

%\cite{Tobe:2003yj}
\bibitem{Tobe:2003yj} 
  K.~Tobe and J.~D.~Wells,
  %``Gravity assisted exact unification in minimal supersymmetric SU(5) and its gaugino mass spectrum,''
  Phys.\ Lett.\ B {\bf 588}, 99 (2004).
%  doi:10.1016/j.physletb.2004.02.072
%  [hep-ph/0312159].
  %%CITATION = doi:10.1016/j.physletb.2004.02.072;%%
  %15 citations counted in INSPIRE as of 26 Apr 2016

%\cite{Hisano:1992mh}
\bibitem{Hisano:1992mh} 
  J.~Hisano, H.~Murayama and T.~Yanagida,
  %``Probing GUT scale mass spectrum through precision measurements on the weak scale parameters,''
  Phys.\ Rev.\ Lett.\  {\bf 69}, 1014 (1992);
%  doi:10.1103/PhysRevLett.69.1014
  %%CITATION = doi:10.1103/PhysRevLett.69.1014;%%
  %121 citations counted in INSPIRE as of 07 Jun 2016
%\cite{Hisano:2013cqa}
%\bibitem{Hisano:2013cqa} 
  J.~Hisano, T.~Kuwahara and N.~Nagata,
  %``Grand Unification in High-scale Supersymmetry,''
  Phys.\ Lett.\ B {\bf 723}, 324 (2013).
%  doi:10.1016/j.physletb.2013.05.017
%  [arXiv:1304.0343 [hep-ph]].
  %%CITATION = doi:10.1016/j.physletb.2013.05.017;%%
  %25 citations counted in INSPIRE as of 07 Jun 2016


%
\bibitem{g2mssm} S.~A.~R.~Ellis, G.~L.~Kane and B.~Zheng,
  %``Superpartners at LHC and Future Colliders: Predictions from Constrained Compactified M-Theory,''
  JHEP {\bf 1507} (2015) 081.
%

%\cite{Wells:2003tf}
\bibitem{Wells:2003tf} 
  J.~D.~Wells,
  %``Implications of supersymmetry breaking with a little hierarchy between gauginos and scalars,''
%  hep-ph/0306127.
 hep-ph/0306127;
  %%CITATION = HEP-PH/0306127;%%
  %128 citations counted in INSPIRE as of 29 Apr 2016
%\cite{ArkaniHamed:2004fb}
%\bibitem{ArkaniHamed:2004fb} 
  N.~Arkani-Hamed and S.~Dimopoulos,
  %``Supersymmetric unification without low energy supersymmetry and
	%signatures for fine-tuning at the LHC,''
  {JHEP {\bf 0506}, 073 (2005)};
%[\href{http://arxiv.org/abs/hep-th/0405159}{{\tt
%  hep-th/0405159}}];
%  doi:10.1088/1126-6708/2005/06/073
%  [hep-th/0405159].
  %%CITATION = doi:10.1088/1126-6708/2005/06/073;%%
  %894 citations counted in INSPIRE as of 29 Apr 2016
%
%\cite{Giudice:2004tc}
%\bibitem{Giudice:2004tc} 
  G.~F.~Giudice and A.~Romanino,
  %``Split supersymmetry,''
%\href{http://dx.doi.org/10.1016/j.nuclphysb.2004.11.048}
  {Nucl.\ Phys.\ B {\bf 699}, 65 (2004)};
  Erratum: [{Nucl.\ Phys.\ B {\bf 706}, 487 (2005)}];
%  doi:10.1016/j.nuclphysb.2004.11.048, 10.1016/j.nuclphysb.2004.08.001
%  [hep-ph/0406088].
  %%CITATION = doi:10.1016/j.nuclphysb.2004.11.048, 10.1016/j.nuclphysb.2004.08.001;%%
  %681 citations counted in INSPIRE as of 29 Apr 2016
%\cite{ArkaniHamed:2004yi}
%\bibitem{ArkaniHamed:2004yi} 
  N.~Arkani-Hamed, S.~Dimopoulos, G.~F.~Giudice and A.~Romanino,
  %``Aspects of split supersymmetry,''
%\href{http://dx.doi.org/10.1016/j.nuclphysb.2004.12.026}
  {Nucl.\ Phys.\ B {\bf 709}, 3 (2005)};
%  [\href{http://arxiv.org/abs/hep-ph/0409232}{{\tt hep-ph/0409232}}];
 % doi:10.1016/j.nuclphysb.2004.12.026
 % [hep-ph/0409232].
  %%CITATION = doi:10.1016/j.nuclphysb.2004.12.026;%%
  %506 citations counted in INSPIRE as of 29 Apr 2016
%\cite{Wells:2004di}
%\bibitem{Wells:2004di} 
  J.~D.~Wells,
  %``PeV-scale supersymmetry,''
%\href{http://dx.doi.org/10.1103/PhysRevD.71.015013}
  {Phys.\ Rev.\ D {\bf 71}, 015013 (2005)}.
%[\href{http://arxiv.org/abs/hep-ph/0411041}{{\tt hep-ph/0411041}}].
 % doi:10.1103/PhysRevD.71.015013
 % [hep-ph/0411041].
  %%CITATION = doi:10.1103/PhysRevD.71.015013;%%
  %214 citations counted in INSPIRE as of 29 Apr 2016

%\cite{Giudice:2011cg}
\bibitem{Giudice:2011cg} 
  G.~F.~Giudice and A.~Strumia,
  %``Probing High-Scale and Split Supersymmetry with Higgs Mass Measurements,''
  Nucl.\ Phys.\ B {\bf 858}, 63 (2012);
%  doi:10.1016/j.nuclphysb.2012.01.001
%  [arXiv:1108.6077 [hep-ph]].
  %%CITATION = doi:10.1016/j.nuclphysb.2012.01.001;%%
  %193 citations counted in INSPIRE as of 07 Jun 2016
%\cite{Hall:2011jd}
%\bibitem{Hall:2011jd} 
  L.~J.~Hall and Y.~Nomura,
  %``Spread Supersymmetry,''
%\href{http://dx.doi.org/10.1007/JHEP01(2012)082}
  {JHEP {\bf 1201}, 082 (2012)};
%[\href{http://arxiv.org/abs/1111.4519}{{\tt arXiv:1111.4519}}];
%  doi:10.1007/JHEP01(2012)082
%  [arXiv:1111.4519 [hep-ph]].
  %%CITATION = doi:10.1007/JHEP01(2012)082;%%
  %106 citations counted in INSPIRE as of 29 Apr 2016
%\cite{Ibe:2011aa}
%\bibitem{Ibe:2011aa} 
  M.~Ibe and T.~T.~Yanagida,
  %``The Lightest Higgs Boson Mass in Pure Gravity Mediation Model,''
%\href{http://dx.doi.org/10.1016/j.physletb.2012.02.034}
  {Phys.\ Lett.\ B {\bf 709}, 374 (2012)};
%  [\href{http://arxiv.org/abs/1112.2462}{{\tt arXiv:1112.2462}}];
%  doi:10.1016/j.physletb.2012.02.034
%  [arXiv:1112.2462 [hep-ph]].
  %%CITATION = doi:10.1016/j.physletb.2012.02.034;%%
  %158 citations counted in INSPIRE as of 29 Apr 2016
%\cite{Ibe:2012hu}
%\bibitem{Ibe:2012hu} 
  M.~Ibe, S.~Matsumoto and T.~T.~Yanagida,
  %``Pure Gravity Mediation with m_{3/2} = 10-100TeV,''
%\href{http://dx.doi.org/10.1103/PhysRevD.85.095011}
  {Phys.\ Rev.\ D {\bf 85}, 095011 (2012)};
%  [\href{http://arxiv.org/abs/1202.2253}{{\tt arXiv:1202.2253}}];
%  doi:10.1103/PhysRevD.85.095011
%  [arXiv:1202.2253 [hep-ph]].
  %%CITATION = doi:10.1103/PhysRevD.85.095011;%%
  %138 citations counted in INSPIRE as of 29 Apr 2016
%\cite{Arvanitaki:2012ps}
%\bibitem{Arvanitaki:2012ps} 
  A.~Arvanitaki, N.~Craig, S.~Dimopoulos and G.~Villadoro,
  %``Mini-Split,''
%\href{http://dx.doi.org/10.1007/JHEP02(2013)126}
  {JHEP {\bf 1302}, 126 (2013)};
%[\href{http://arxiv.org/abs/1210.0555}{{\tt
%  arXiv:1210.0555}}];
 % doi:10.1007/JHEP02(2013)126
  %%CITATION = doi:10.1007/JHEP02(2013)126;%%
  %174 citations counted in INSPIRE as of 29 Apr 2016
%\cite{Hall:2012zp}
%\bibitem{Hall:2012zp} 
  L.~J.~Hall, Y.~Nomura and S.~Shirai,
  %``Spread Supersymmetry with Wino LSP: Gluino and Dark Matter
	%Signals,''
%\href{http://dx.doi.org/10.1007/JHEP01(2013)036}
  {JHEP {\bf 1301}, 036 (2013)};
%[\href{http://arxiv.org/abs/1210.2395}{{\tt arXiv:1210.2395}}];
%  doi:10.1007/JHEP01(2013)036
%  [arXiv:1210.2395 [hep-ph]].
  %%CITATION = doi:10.1007/JHEP01(2013)036;%%
  %91 citations counted in INSPIRE as of 29 Apr 2016
%\cite{ArkaniHamed:2012gw}
%\bibitem{ArkaniHamed:2012gw} 
  N.~Arkani-Hamed, A.~Gupta, D.~E.~Kaplan, N.~Weiner and T.~Zorawski,
  %``Simply Unnatural Supersymmetry,''
%  \href{http://arxiv.org/abs/1212.6971}{{\tt arXiv:1212.6971}};
  arXiv:1212.6971 [hep-ph];
  %%CITATION = ARXIV:1212.6971;%%
  %157 citations counted in INSPIRE as of 29 Apr 2016
%\cite{Evans:2013lpa}
%\bibitem{Evans:2013lpa} 
  J.~L.~Evans, M.~Ibe, K.~A.~Olive and T.~T.~Yanagida,
  %``Universality in Pure Gravity Mediation,''
%\href{http://dx.doi.org/10.1140/epjc/s10052-013-2468-9}
  {Eur.\ Phys.\ J.\ C {\bf 73}, 2468 (2013)};
% [\href{http://arxiv.org/abs/1302.5346}{{\tt arXiv:1302.5346}}].
 % doi:10.1140/epjc/s10052-013-2468-9
%  [arXiv:1302.5346 [hep-ph]].
  %%CITATION = doi:10.1140/epjc/s10052-013-2468-9;%%
  %35 citations counted in INSPIRE as of 29 Apr 2016
%\cite{Bagnaschi:2014rsa}
%\bibitem{Bagnaschi:2014rsa} 
  E.~Bagnaschi, G.~F.~Giudice, P.~Slavich and A.~Strumia,
  %``Higgs Mass and Unnatural Supersymmetry,''
  JHEP {\bf 1409}, 092 (2014).
%  doi:10.1007/JHEP09(2014)092
%  [arXiv:1407.4081 [hep-ph]].
  %%CITATION = doi:10.1007/JHEP09(2014)092;%%
  %51 citations counted in INSPIRE as of 07 Jun 2016

%\cite{Randall:1998uk}
\bibitem{Randall:1998uk} 
  L.~Randall and R.~Sundrum,
  %``Out of this world supersymmetry breaking,''
%\href{http://dx.doi.org/10.1016/S0550-3213(99)00359-4}
  {Nucl.\ Phys.\ B {\bf 557}, 79 (1999)};
%  [\href{http://arxiv.org/abs/hep-th/9810155}{{\tt hep-th/9810155}}].
%  doi:10.1016/S0550-3213(99)00359-4
%  [hep-th/9810155].
  %%CITATION = doi:10.1016/S0550-3213(99)00359-4;%%
  %1492 citations counted in INSPIRE as of 29 Apr 2016
%
%\cite{Giudice:1998xp}
%\bibitem{Giudice:1998xp} 
  G.~F.~Giudice, M.~A.~Luty, H.~Murayama and R.~Rattazzi,
  %``Gaugino mass without singlets,''
%\href{http://dx.doi.org/10.1088/1126-6708/1998/12/027}
  {JHEP {\bf 9812}, 027 (1998)}.
%  [\href{http://arxiv.org/abs/hep-ph/9810442}{{\tt hep-ph/9810442}}].
%  doi:10.1088/1126-6708/1998/12/027
%  [hep-ph/9810442].
  %%CITATION = doi:10.1088/1126-6708/1998/12/027;%%
  %1192 citations counted in INSPIRE as of 29 Apr 2016


\end{thebibliography}
\end{document}